\documentclass[prd,onecolumn,preprintnumbers,nofootinbib]{revtex4}
\usepackage{graphicx}

\usepackage[plainpages=false, colorlinks=true, anchorcolor=blue, linkcolor=blue, citecolor=blue, bookmarks=false]{hyperref}
\usepackage{color}
\usepackage{float}
\usepackage{amsfonts,amsmath,amssymb}
\usepackage{natbib}
\usepackage{array}
\begin{document}
\newcolumntype{P}[1]{>{\centering\arraybackslash}p{#1}}
\pdfoutput=1
\newcommand{\jcap}{JCAP}
\newcommand{\araa}{Annual Review of Astron. and Astrophys.}
\newcommand{\rthis}[1]{\textcolor{black}{#1}}
\newcommand{\aj}{Astron. J. }
\newcommand{\mnras}{MNRAS}
\newcommand{\apjl}{Astrophys. J. Lett.}
\newcommand{\apjs}{Astrophys. J. Suppl. Ser.}
\newcommand{\actaa}{Acta Astronomica} 
\newcommand{\aap}{Astron. \& Astrophys.}
\newcommand{\aapr}{Astronomy and Astrophysics Review }
\renewcommand{\arraystretch}{2.5}
\title{Search for spatial coincidence between magnetars  and  IceCube detected neutrinos}
\author{Fathima \surname{Shifa M}}
\altaffiliation{Current Address: Department of Physics, South Dakota School of Mines and Technology, Rapid City, SD 57701, USA}
\author{Shantanu \surname{Desai}}
\altaffiliation{E-mail: shntn05@gmail.com}
\affiliation{Department  of Physics, IIT Hyderabad,  Kandi, Telangana-502284, India}

\begin{abstract}
We implement a search for  spatial coincidence between high energy neutrinos detected by the IceCube neutrino detector (using the publicly available 10-year muon track data) and  37 magnetars, including six extragalactic sources.  We use the 
unbinned maximum likelihood method for our analysis.  We do not find any such spatial association between any of  the known magnetars and  IceCube-detected neutrinos. Therefore, we conclude that none of the known galactic or extragalactic magnetars  contribute to the diffuse neutrino flux observed in  IceCube.  A stacked analysis also does not show a statistically significant excess.
\end{abstract}

\maketitle
\section{Introduction}
The origin of the dominant component of the IceCube  diffuse neutrino flux observed in the TeV-PeV energy  range~\cite{IceCubescience} is still  unknown~\cite{Halzen23}. 
Although, evidence for neutrino emission from a few selected  point sources such as  NGC 1068, TXS 0506+056, NGC 4151, and PKS 1424+240 has been found, most of the  IceCube neutrinos cannot be attributed to any particular astrophysical sources~\cite{IceCubedata}. 
 Therefore, searches for spatial coincidence with a large number of extragalactic sources have been done. Some examples of such  extra-galactic sources  include  Active Galactic Nuclei (AGNs), Gamma-Ray bursts (GRBs)~\cite{Kamionkowski,Hooper,LuoZhang,Smith21,Li22,Icecube10,IceCubeGRB,IceCubeAGN,IceCubeblazars}, Fast Radio bursts (FRBs) ~\cite{Zhang21,Desai23,IceCubeFRB}, tidal disruption events~\cite{LiTDE}, Fermi-LAT point sources~\cite{Li22},  galaxy mergers~\cite{Laha}, galaxies from extra-galactic surveys such as   2MASS~\cite{IceCube2MASS} and WISE-2MASS~\cite{Fang24}, etc.

At the same time, a galactic contribution to the IceCube diffuse neutrino flux cannot be ruled out,  and has been estimated to be up to 20\%~\cite{Palladino,Troitsky}. 
Therefore, searches for coincidences with a plethora of galactic sources such as pulsars, supernova remnants, X-ray binaries,  TeV gamma-ray sources,  open clusters, red dwarfs, etc. have also been carried out~\cite{Lunardini,Marfatia,IceCubeXRB,IceCubePWN,Kovalev22,LHAASO,Pasumarti,Pasumarti2,Rott,Shifa}.

In this work, we search for high-energy neutrinos from magnetars. Magnetars are a particular type of  neutron stars with extremely  strong magnetic fields, with values up to $10^{15}$G and X-ray luminosities between $10^{31}-10^{34}$ ergs/sec~\cite{Woods,Mereghetti,Kaspirev,Espositorev,Rea25}. They have also been seen at optical and radio wavelengths~\cite{Rea25}, but not at   gamma-ray energies~\cite{Vyaas}.
    The magnetar X-ray  emission stems  from the decay and instability of their ultra-strong magnetic fields~\cite{DuncanThompson,Paczynski,ThompsonDuncan}.
    Observationally, they are manifested as soft gamma-ray repeaters (SGRs) and anomalous X-ray pulsars (AXP)~\cite{Mereghetti}. Although historically most magnetars were known to be galactic~\cite{Woods}, we  have now detected extragalactic magnetar flares from a number of sources.
Several theoretical models for neutrino emission from magnetars have been proposed during their quiescence phase~\cite{Zhang03,Luo05,Murase09,Dey16} as well as during their outbursts~\cite{Halzen05,Ioka05} (see ~\cite{Negro24} for a recent review).
In the first proposed  model for neutrino emission from magnetars~\cite{Zhang03},  it was asserted that young magnetars with opposite orientations of spin moments and magnetic fields accelerate cosmic rays. Neutrino emission then occurs from the  decay of pions produced through photo-meson interactions. The two main sources of energy that power a magnetar are  its  rotational energy loss (which accelerates the protons) and magnetic field decay (which  provides the target column density of photons for photo-meson production). Then in another work, ~\citet{Dey16} estimated the flux of PeV neutrinos from photomeson interactions in the magnetic polar caps, after considering the enhanced  radiative background due to photon splitting in very strong magnetic fields.  
    
    Magnetars show  different types of transient emission, such as short bursts, intermediate flares and giant flares~\cite{Rea11,Rea25} with time scales
    ranging from milliseconds to years.
    The short bursts last for about 0.1-0.2 seconds with peak luminosities of about $10^{38}-10^{40}$ erg. The intermediate flares last from several seconds to years
    with luminosities between $10^{41}-10^{42}$ ergs. The giant flares are the most energetic with luminosities between $10^{43}-10^{45}$ ergs/sec. However, these giant flares for galactic magnetars have only been observed for three  sources before the start of IceCube data taking~\cite{Esposito}. 
    Furthermore, 
    the exact time of the onset of outbursts for magnetars is observationally difficult to determine~\cite{Rea11}. Observationally, it has also  been found  that the  luminosity of many SGRs during quiescence is enhanced for a long period of time~\cite{Woods03}. 
     Therefore, it has also been argued that the neutrino flux during the post-burst phase could be greater than during quiescence due to more facilitated conditions~\cite{Zhang03}.
    Therefore, it is important to look for neutrinos during both their flaring and quiescent states. For all these reasons, we carried out a time-integrated search for neutrino emission from magnetars using the entire IceCube public data. 
    
    The first ever  search for neutrinos from magnetars with a 1000~$\rm{m^2}$ detector was done using Super-Kamiokande~\cite{Desai04,sknuastro}. In that analysis,  a search for neutrinos in temporal coincidence with four SGR bursts using the upward-going muon sample observed in Super-K was carried out. Although two neutrinos were seen in spatial coincidence within $\pm 5^{\circ}$ and within a day after the SGR outburst, the observed $p$-value is consistent with background, once you take into account the look-elsewhere effect~\cite{sknuastro}.  Subsequently, a search during both the flaring and quiescent phases was done using the highest-energy neutrino subset  in Super-Kamiokande, but no signal was detected~\cite{skshowering}. A proof of principle study of the sensitivity of IceCube to detect neutrinos from galactic magnetars using 14 years of IceCube data has also been carried out~\cite{Icecubemagnetar,IceCubemagnetar2}.


In this work, we search for neutrino emission from both galactic and extragalactic magnetars using the publicly available IceCube dataset. We follow the same prescription as in our previous works~\cite{Pasumarti,Shifa}. This manuscript is structured as follows. The neutrino data set used for the analysis is discussed in Section~\ref{sec:dataset}. The analysis and results for single source analysis are discussed in Section~\ref{sec:analysis} and Section~\ref{sec:results}, respectively.  The results of the stacked analysis are discussed in Sect.~\ref{sec:stacked}. We conclude in Section~\ref{sec:conclusions}.

\section{Dataset}
\label{sec:dataset}
For this analysis, we have used  neutrinos from the IceCube public 10-year muon track data~\cite{IceCubedata}. This data set consists of 1,134,431 neutrinos, which were collected between April 2008 (IC-40) and July 2018 (IC86-VII) from different phases of the experiment, each having a different lifetime. For each neutrino, the data set consists of right ascension (RA), declination, uncertainty in the direction of the track, and reconstructed muon energy. Note that for our analysis, we have used the augmented data set analyzed in ~\cite{Beacom}, which removes some duplicates from the IceCube dataset~\footnote{This dataset is available at \url{https://github.com/beizhouphys/IceCube_data_2008--2018_double_counting_corrected}}. The list of galactic  magnetars used for this work has been obtained  from the McGill Online Magnetar Catalog ~\cite{2014ApJS..212....6O},\footnote{This dataset is available at \url{http://www.physics.mcgill.ca/~pulsar/magnetar/main.html}} and includes 31 magnetars, of which  16 are SGRs (12 confirmed candidates), and  14 are AXPs (12 confirmed candidates). One of the magnetar candidates includes PSR J1846-0258,  which has been classified as a young, rotation-powered pulsar, but has also undergone a  magnetar-like outburst in 2006~\cite{Gavri08}.
In addition to these galactic magnetars, we also considered six extragalactic magnetars, for  which flares were  detected in high-energy gamma-rays. These include   GRB 200415A, GRB 2311115A, GRB 051103, GRB 070201, GRB 070222, and GRB 180128A.

\section{Analysis}
\label{sec:analysis}
For our analysis, we use the unbinned maximum likelihood ratio method~\cite{Montaruli}. We consider neutrinos within a declination of $\pm 5^{\circ}$ of the magnetars. For a dataset containing $N$ \rthis{neutrino} events, if $n_s$ is the total number of signal events  attributed to the magnetars, then the likelihood is given by:
\begin{equation}
\mathcal{L} (n_s) = \prod_{i=1}^N \left[\frac{n_s}{N} S_i + (1-\frac{n_s}{N}) B_i\right], 
\label{eq1}
\end{equation}
where $S_i$ is the signal probability density function (PDF) and $B_i$ is the background PDF. Here, the signal PDF is given by:
\begin{equation}
S_i = \frac{1}{2\pi\sigma_i^2}e^{-(|\theta_i-\theta_j|)^2/2\sigma_i^2}, 
\label{eq:2}
\end{equation}
where $|\theta_i-\theta_j|$ is the angular distance between the  neutrino and the magnetar (indexed by $j$); $\sigma_i$ is the angular uncertainty in the neutrino position expressed in radians. Similarly to other works which have analyzed the IceCube public data~\cite{Kamionkowski,Pasumarti,Li22}, we have estimated the background PDF from the data. We assume that the background neutrinos are uniformly distributed across the region of interest and do not have any dependence on the right ascension. The background PDF can therefore be obtained from  the solid angle within the declination band  ($\delta$) of  $\pm 5^{\circ}$ around each magnetar, and is given as follows~\cite{Pasumarti,Li22,Montaruli}:
\begin{equation}
B_i=\frac{1}{\Omega_{\delta \pm 5^{\circ}}}
\end{equation}
Note that 99\% of neutrino events in the 10 year muon track data have angular reconstruction error $<5^{\circ}$.  Therefore, in order to be conservative we chose a declination band of $ \pm 5^{\circ}$ so as to account for the angular reconstruction error of most of the detected neutrino-induced muons. In one of our previous works involving searches for neutrinos from pulsars, we have also confirmed that the results do not drastically change with other similar choices of declination bands such as $3^{\circ}$, $4^{\circ}$~\cite{Pasumarti2}.

Now to ascertain the significance of the signal, we define the  test statistics (TS) as follows:
\begin{equation}
TS (n_s) = 2 \log \frac{\mathcal{L} (\hat{n}_s)}{\mathcal{L} (0)},
\label{eq:ts}
\end{equation}
where $\hat{n}_s$ corresponds to the value of $n_s$ which maximizes  $\mathcal{L} (n_s)$.
The probability distribution function of TS is given by the superposition of  $\delta$ function  and a $\chi^2$ distribution,  due to the fact that the fit parameters $n_s \geq  0$ is bounded~\cite{Wolf,Mattox96,Cowan11,Chernoff}.
Therefore, for the null hypothesis, TS  behaves  asymptotically like a half-$\chi^2$ distribution  for one degree of freedom ($\frac{1}{2} \delta (x)+ \frac{1}{2} \chi^2$) due to a modification of Wilk's theorem~\cite{Chernoff,Mattox96,Cowan11}.
To evaluate the significance, the $p$-value can be  obtained from  half the value of the survival function (or the complementary cumulative distribution function) for the  chi-square distribution with one  degree of freedom, evaluated at TS~\cite{Chernoff,Mattox96,Cowan11}.
The detection significance (or $Z$-score) can therefore be approximated  as $\sqrt{TS}$~\cite{Mattox96,Cowan11}. For a statistically significant detection corresponding to $>5\sigma$ detection, TS must be $> 25$. 
We shall also independently check the distribution of TS for the null hypothesis in Section~\ref{sec:results}.

\section{Results}
\label{sec:results}
For each of our magnetars, we calculate the best fit $n_s$ that maximizes TS, according to Eq.~\ref{eq:ts}.   These TS values, along with the best-fit value of $n_s$, can be found in Table~\ref{table1}. As we can see, none of the magnetars show a TS value $> 25$. The largest TS value is seen for 3XMM J185246.6+003317,  with TS of 3.34, which corresponds to a significance of around $1.8\sigma$ (based on the modification of Wilks' theorem as discussed in the previous section).

\begin{figure}[htbp]
    \centering
    \includegraphics[width=0.8\textwidth]{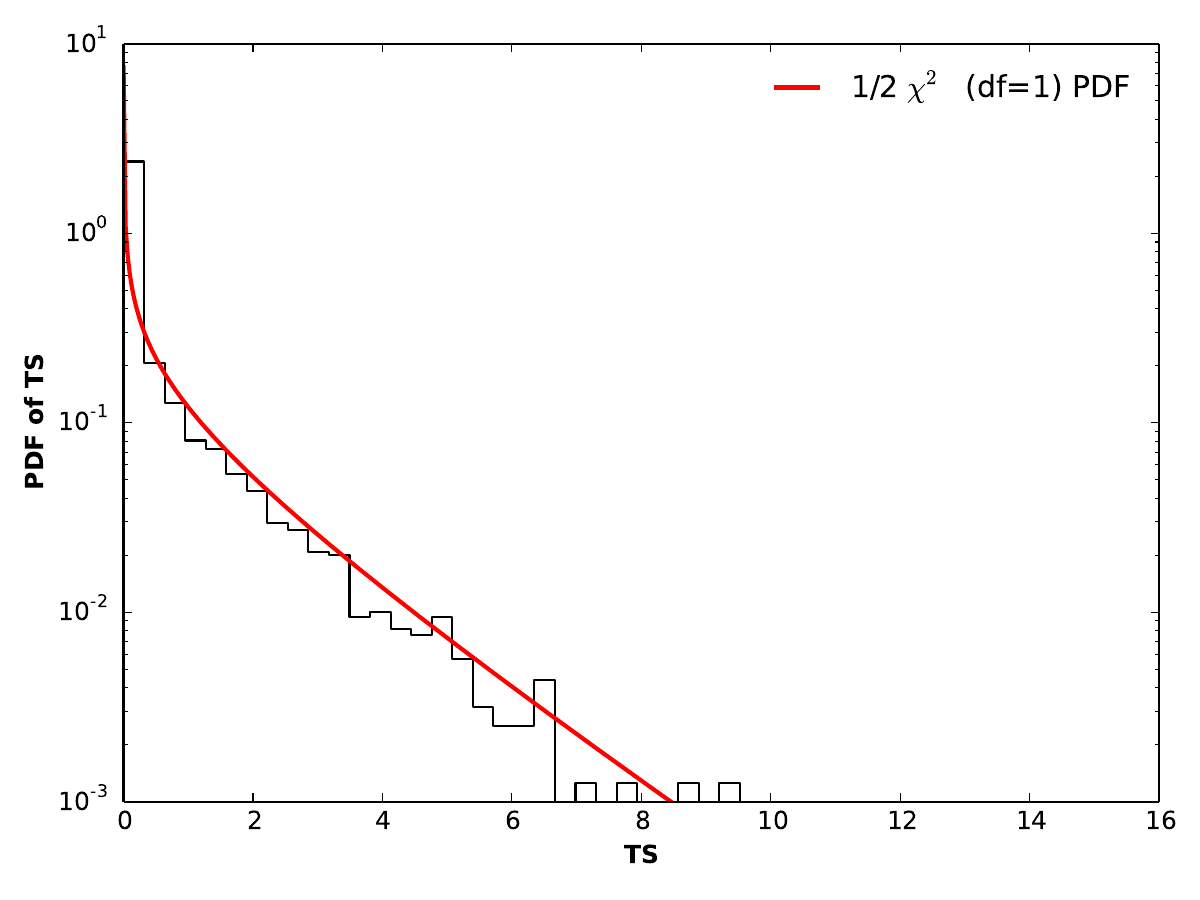}
    \caption{\label{fig:ts_histogram} Distribution of TS for null hypothesis using 5000 randomly selected positions uniformly distributed between RA of 0 and $360^{\circ}$, and $\sin(\delta)$ between -1 and 1. The red line  curve corresponds to  half the PDF of $\chi^2$ distribution with one degree of freedom and provides a good fit to the PDF of TS distribution for TS$>0$.}
\end{figure}

To independently evaluate the $p$-value for the magnetar with the highest TS value, we plot the distribution of TS for the null hypothesis of no signal.
For this purpose, similar  to ~\cite{Kamionkowski}, we evaluated TS at  5000 locations on the celestial sphere with RA uniformly distributed between 0 and $360^{\circ}$,  and $\sin (\delta)$ uniformly distributed between $-1.0$ and $1.0$.
We then evaluate TS for each of these random locations in the same way as for each of these magnetars. The probability distribution function (PDF) of  TS for each of these random locations can be found in Fig.~\ref{fig:ts_histogram} and constitutes the null hypothesis. The red curve corresponds to half the PDF of $\chi^2$ distribution for one degree of freedom. Therefore,  TS follows a half $\chi^2$ distribution for one degree of freedom  as  expected~\cite{Mattox96,Cowan11}.
From this distribution, the $p-$value can also be found non-parametrically,  by calculating the ratio of number of simulated locations with TS values greater than that observed for a given magnetar to the total number of simulated TS values. If we consider the maximum TS value of 3.34 (for XMM J185246.6 + 003317), we find a total of 129 simulated locations with \rthis{TS$>3.34$}, which corresponds to a $p$-value of 0.026, which in turn corresponds to an approximate  significance of 1.9$\sigma$~\cite{Cowan11,Ganguly17}, which roughly agrees with the value of 1.8$\sigma$ obtained from $\sqrt{TS}$, which we had estimated earlier. The post-trials $p$-value after accounting for the look-elsewhere effect is given by $1 - (1-0.026)^{37}$, which is equal to 0.6 and hence is consistent with background.

Therefore, the observed signal events are consistent with background, and there is no evidence for any spatial association between our catalog of magnetars and IceCube neutrinos, which are observed  as muon tracks. Therefore, none of the known magnetars contribute towards the diffuse IceCube neutrino signal. Consequently, we  calculate the 95\% confidence level (c.l.) upper  limits on the observed signal events  by calculating the value of $\hat{n}_s$, for which $TS-TS_{max}=-2.71$, since the asymptotic values of TS obey a half-chisquare distribution.
\rthis{In other words $0.5+0.5 \int_{0}^{2.71} \chi^2_1 (\xi) d\xi=0.95$, where 
$\chi^2_1 (\xi)$ represents the $\chi^2$ for one degree of freedom. }
In order to compare to theoretical models, we first obtain the differential neutrino flux limits. 
In order to do this, we assume that the magnetar neutrino spectrum can be characterized as a function of neutrino energy ($E_{\nu}$) by a power law with spectral index $\Gamma$.
\begin{equation}
\Phi_{\nu} (E_{\nu})=\phi_0 \left(E_{\nu}/100~\text{TeV}\right)^{\Gamma}
\label{eq:phinu}
\end{equation}
where $\phi_0$ is the flux normalization at 100 TeV
For this purpose we calculate the effective acceptance across all the $k$ seasons for a magnetar with declination ($\delta_j$) which is given by
\begin{equation}
A_{acc}= \sum_{i=1}^k T_k \times \int A_{eff}(E_{\nu}, \delta_j) \Phi_{\nu} (E_{\nu}) dE_{\nu},
\label{eq:acc}
\end{equation}
where $\phi_{\nu} (E_{\nu})$ is defined in Eq.~\ref{eq:phinu},  To evaluate the acceptance, we have assumed the  spectral index ($\Gamma=-2.53$), similar to ~\cite{Pasumarti2}. In Eq~\ref{eq:acc},
$T_k$ is the livetime of each season and the sum is over all the IceCube seasons. The IceCube effective area ($A_{eff}(E_{\nu}, \delta_j)$) is a function of the neutrino energy in addition to the $\delta_j$.  For each magnetar, we then calculated the neutrino flux limits at $E_{\nu}=100$ TeV by dividing the 95\% upper limit on the number of signal events by $A_{acc}$. These flux limits can be trivially computed for any other neutrino energy using Eq.~\ref{eq:phinu}.
These neutrino flux limits for the spectral index of  $\Gamma=-2.53$ can be found in the last column in Table ~\ref{table1}. 

The corresponding flux for neutrino-induced muons produced from magnetars where spin and magnetic moments point in opposite directions has been estimated for four objects,  and found to be $\mathcal{O}$ (0.01-10) $km^{-2} yr^{-1}$~\cite{Zhang03} with the maximum expected flux for SGR 1900+14, equal to 13 $\rm{km^{-2} yr^{-1}}$. The corresponding fluxes of PeV energy neutrino-induced muons produced from
acceleration in the magnetic polar caps are even lower, viz. $2.5 \times 10^{-4}$ $\rm{ km^{-2} yr^{-1}}$~\cite{Dey16}.

In order to compare our neutrino flux limit to theoretical predictions,  we convert our estimated neutrino flux limits to upward  muon flux limits ($\Phi_{\mu}$) using the same prescription as in ~\cite{skshowering}:
\begin{equation}
\Phi_{\mu}= \int_{E_{th}}^{\infty} \Phi_{\nu} (E_{\nu}) P(\nu \rightarrow \mu) d E_{\nu},
\end{equation}
where $P(\nu \rightarrow \mu)$  is the muon to neutrino conversion probability obtained from $P(\nu \rightarrow \mu) \approx 1.3 \times 10^{-6} (E_{\nu}/TeV) $; $E_{th}$  is the muon energy threshold which is assumed to be 200 GeV.  Using this we obtain the 95\% c.l.  muon flux limits for SGR 1900+14, SGR 0526-66, 1E 1048-66, and SGR 1806-20 to be $\sim 10^6\rm{km^{-2}yr^{-1}}$. 
Therefore, our muon flux limits are not stringent enough to constrain models of neutrino emission from rotating magnetars with spin and magnetic moments oppositely aligned~\cite{Zhang03} or the mechanism proposed in ~\cite{Dey16}. 

\section{Stacked Analysis}
\label{sec:stacked}
We now perform a stacked search analysis for the magnetars, following the same methodology as in ~\cite{Pasumarti2}. We quickly summarize the procedure followed  (using the same notation as ~\cite{Pasumarti2}), and highlight some differences with respect to single-source analysis. 

The background PDF ($B_i$) for stacked analysis is determined by the solid angle within  $5^{\circ}$ declination band   around  each event $i$ ($\Omega_{\delta_i \pm 5^{\circ}}$), and scale by the fraction of events within this declination band:
\begin{equation}
    B_i=\frac{\mathcal{N}_i}{N\Omega_{\delta_i \pm 5^{\circ}}}, 
    \label{eq:Bistacked}
\end{equation}
where $\mathcal{N}_i$ is the number of events within the $\pm 5^\circ$ declination band of event $i$ and $N$ is the total number of neutrino events in the data set.

The signal PDF ($S_i$) is a weighted average of the signal PDFs ($S_{ij}$) of all the sources:
\begin{equation}
    S_i = \dfrac{\sum_j \omega_{acc, j}\omega_{model, j} S_{ij}}{\sum_j \omega_{acc, j}\omega_{model, j}}, 
    \label{eq:sistacked}
\end{equation}
where the sum $j$ extends over all the 37 magnetars in the catalog,
$S_{ij}$ is given by the same equation as in Eq~\ref{eq:2}. In Eq~\ref{eq:sistacked}, $\omega_{model, j}$ corresponds to the signal weight used for a  given magnetar. Since there is not much guidance from the theoretical models, we choose $\omega_{model, j}=1$ for our analysis. However, it  is straightforward to extend this analysis to other signal weighting factors similar to ~\cite{Pasumarti2}. Finally, $\omega_{acc, j}$
is defined as follows:
\begin{equation}
    \omega_{acc,j} = \sum_i^k T_k   \times \int A_{eff}(E_{\nu}, \delta_j)E_{\nu}^{\Gamma} dE_{\nu} ,
    \label{eq:weights}
\end{equation}
All terms in the above expression have the same meaning as in Eq.~\ref{eq:acc}. For the stacked analysis, we  choose three spectral indices: $\Gamma=-2.0,-2.53,-3.0$.

The expected number of signal events from the source  (${n}_s$)  is then given by~\cite{Zhang21}:
\begin{equation}
    {n}_s = \sum_j {n}_{sj}
    \label{eq:stackedns}
\end{equation}
where
\begin{equation}
    n_{sj} =  \sum_i^k T_k   \times \int A_{eff}(E_{\nu}, \delta_j)\frac{dF}{dE_{\nu}}dE_{\nu},
    \label{eq:ns_HAT-no_source}
\end{equation}
where $n_{sj}$ is the number of signal events coming from magnetar $j$ and $\dfrac{dF}{dE_{\nu}}$ is the expected neutrino spectrum from the source, which is defined in Eq.~\ref{eq:phinu}.

\begin{figure}[htbp]
    \centering
    \includegraphics[width=0.7\textwidth]{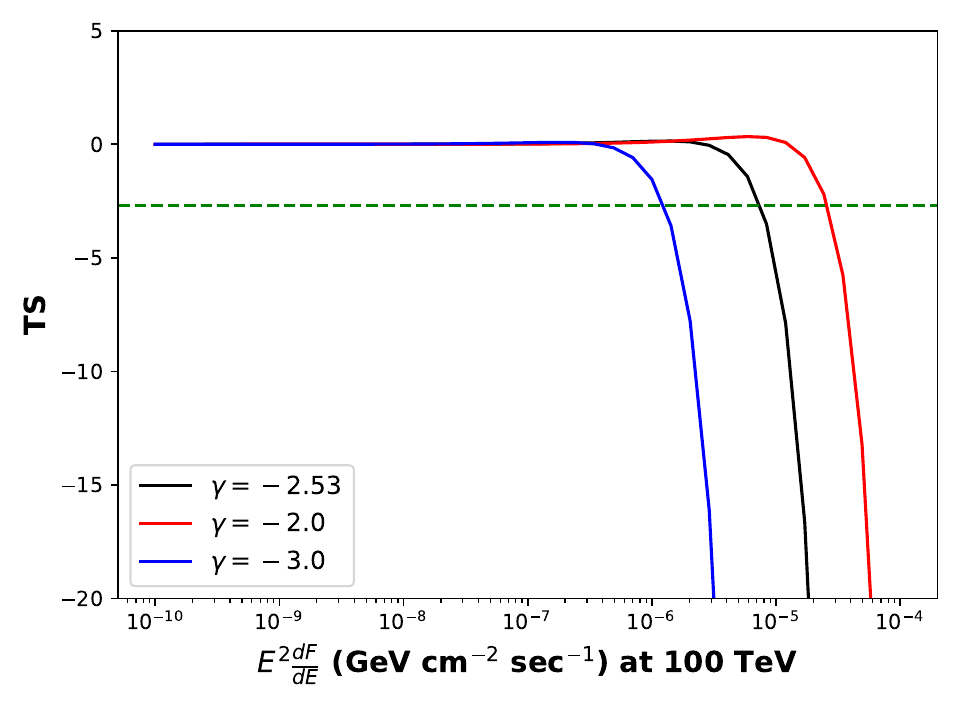}
    \caption{\label{fig:w2}
    Plot of TS  as a function of the total differential  neutrino flux for 37 magnetars  using the IceCube 10-year muon-track data. In this analysis, we considered  $\omega_{model}=1$ and $\Gamma=-2.53,-2,-3$. The dashed horizontal line corresponds to TS=-2.7, which can be used to obtain the 95\% c.l. upper limit.
    The stacked 95\% c.l. upper limit on the differential neutrino flux is given by  $1.2 \times  10^{-6} (\gamma=-3.0), 7.4 \times  10^{-6} (\gamma=-2.53)$, and  $2.6 \times 10^{-5} (\gamma=-2.0)$    $\rm{GeV^{-1}~cm^{-2}~sec^{-1}}$. }
\end{figure}

In order to obtain the limits (or the central  estimates in case of a detection) of the differential neutrino flux, we obtain $n_s$   calculate  TS as a function of $n_s$
over a broad range of neutrino energies. This would automatically help us estimate $\hat{n_s}$ which maximizes the likelihood.  We recap  this procedure to obtain TS for stacked analysis. We start with a given value of $E^2\dfrac{dF}{dE_{\nu}}$, where $E$ denotes the neutrino energy. We then evaluate $n_s$ using Eq.~\ref{eq:stackedns} and Eq.~\ref{eq:ns_HAT-no_source}, and $S_i$ using Eq.~\ref{eq:sistacked}.
Similarly to a single source analysis, we then plug the expressions for $B_i$ and $S_i$ from Eq.~\ref{eq:Bistacked} and Eq.~\ref{eq:sistacked}, respectively, to evaluate $\mathcal{L} (n_s)$   in Eq.~\ref{eq1}.  TS is then evaluated by using the expression for $\mathcal{L} (n_s)$ in the numerator of Eq.~\ref{eq:ts} instead of  $\mathcal{L} (\hat{n}_s)$. 
Similarly to ~\cite{Pasumarti2}, we then  show the plot for  TS over a large dynamic range of $E^2\dfrac{dF}{dE_{\nu}}$, evaluated at  neutrino energy of 100 TeV. 
This plot of TS can be found in Fig.~\ref{fig:w2} for all three spectral indices considered. We can see that the TS value is close to 0 and then gradually declines for $E^2 \dfrac{dF}{dE_{\nu}} \gtrsim   (10^{-6}-10^{-4}) \times 10^{-6}~\rm{GeV cm^{-2} sec^{-1}}$, depending on the spectral index. The maximum value of TS is equal to 0.3 for $\gamma=-2$ and hence is not statistically significant.
Therefore,   we do not see a statistically significant excess using a stacked analysis. We calculated the 95\% c.l. stacked upper limit on the differential muon neutrino flux by finding the X-intercept corresponding to TS=-3.84, which is equal  to $1.2 \times  10^{-6}, 7.4 \times  10^{-6}$,  and  $2.6 \times 10^{-5}~\rm{GeV~cm^{-2}~s^{-1}}$ for the spectral indices of $-3.0,-2.53,$ and $-2.0$, respectively. These limits need to be multiplied by a factor of three, if we want to convert them into  all-flavor neutrino flux~\cite{Li22}. The IceCube diffuse flux from the galactic plane for the $\Pi^0$ analysis  has been estimated to be $\sim 2.18 \times 10^{-8}~\rm{GeV cm^{-2} s^{-1}}$~\cite{Science} and is about 2-3 orders of magnitude smaller than the flux limit,  which we obtained from our magnetar analysis.  \rthis{Note however that the diffuse flux measurement was obtained using cascade events, which are mostly tau and electron neutrinos, whereas our analysis was done using the muon track events.}

\section{Conclusions}
\label{sec:conclusions}
In this work, we searched for spatial coincidence between IceCube neutrinos and 37 magnetars including two extragalactic magnetars from which giant flares were seen. For our analysis, we used the  IceCube 10 year muon track data observed between 2008-2018.   Here, we analyzed our data using the unbinned maximum likelihood method.  A tabular summary of  our results can be found in Table~\ref{table1}. We do not find a detection significance greater than $5\sigma$ for any of our magnetars and the maximum significance is around 2.3$\sigma$ for 3XMM J185246.6+003317. We also calculate the 95\% c.l. upper limit on the number of observed signal events for each of the magnetars. Hence,  we conclude that none of the known galactic magnetars contributes to the IceCube diffuse neutrino flux. We also calculate the muon flux limit for all our magnetars,  and conclude that they  are  not stringent enough to rule out the theoretical models of neutrino emission proposed for some magnetars  in \cite{Zhang03,Dey16}.  Finally, we also did a stacked analysis using all the 37 magnetars assuming equal model weights. The results from the stacked analysis can be found in Fig.~\ref{fig:w2}. As we can see, there is no statistically significant excess from the stacked analysis.

For a more robust test from individual magnetars, next-generation detectors such as IceCube-Gen2~\cite{IceCubeGen2} would be needed.

\begin{center}
\begin{table}[h!]
\setlength{\tabcolsep}{12pt}
\renewcommand{\arraystretch}{0.9} 
    \begin{tabular}{|l|l|l|l|l|l|l|}
    \hline
    \centering
        \textbf{Magnetars} & \textbf{RA ($^{\circ}$)} & \textbf{Decl. (${^\circ}$)} & $\mathbf{n_s}$  & $\mathbf{TS_{max}}$  &\textbf{Upper limit} & \textbf{ $\nu$ Flux limit} \\ 
        & & & & & & $\rm{(erg^{-1}~cm^{-2}~sec^{-1})}$ \\
        \hline
CXOU J010043.1-721134 & 15.17       & 72.17 &9.438               & 1.05   &    30.5   &$4.6\times  10^{-12}  $  \\  \hline
4U 0142+61            & 26.59       & 61.75  &0.0             & 0.0    &   16.79      &$2.4\times10^{-12}  $\\ \hline
SGR 0418+5729         & 64.64      & 57.53  & 0.961              & 0.01         & 18.2     &$2.5\times10^{-12}$   \\ \hline
SGR 0501+4516         & 75.27       & 45.27  & 0.0            & 0.0 &  11.39  &$1.5\times10^{-12} $     \\ \hline
SGR 0526-66           & 81.50    & 66.05  &3.08             & 0.15     &  20.3 &$2.9\times10^{-12} $    \\\hline
1E 1048.1-5937        & 162.52      & 59.87   & 0.0              & 0.0    &    8.99    &$1.2\times10^{-12} $  \\\hline
1E 1547.0-5408        & 237.72     & 54.29   & 0.954         & 0.008   &      20.5  &$2.7\times10^{-12}$     \\\hline
PSR J1622-4950        & 245.68     & 49.81   & 1.62           & 0.02  &      22.1   &$3.0\times10^{-12}$     \\\hline
SGR 1627-41&  248.96 &47.57 &0.0&0.0 &  12.09 &$1.6\times10^{-12} $  \\ \hline
CXOU J164710.2-455216 & 251.79     & 45.86   &4.75           & 0.17  &      11.49  &$1.5\times10^{-12}$   \\\hline
1RXS J170849.0-400910 & 257.19      & 40.11   & 3.21             & 0.09  &    23.89      &$3.2\times10^{-12} $    \\\hline
CXOU J171405.7-381031 & 258.52      & 38.15   & 0.0               &0.0   &      16.89   &$2.2\times10^{-12} $    \\\hline
SGR J1745-2900        & 266.41      & 28.99   & 2.44             & 0.09 &    16.49   &$2.1\times10^{-12 }$       \\\hline
SGR 1806-20           & 272.16      & 20.38   & 0.0                & 0.0   &       11.99   &$1.5\times10^{-12} $      \\\hline
XTE J1810-197         & 272.46      & 19.70   & 0.0                & 0.0   &      13.69   &$1.7\times10^{-12 }$    \\\hline
Swift J1818.0-1607    & 274.51     & 16.10   & 0.0                & 0.0    &    16.59&$2.1\times10^{-12} $   \\\hline
Swift J1822.3-1606    & 275.57     & 16.05   & 19.62              & 2.61   &      43.8   &$5.6\times10^{-14 }$      \\\hline
SGR 1833-0832         & 278.43     & 8.51    & 0.0                & 0.0    &      17.19    &$2.2\times10^{-12} $   \\\hline
Swift J1834.9-0846    & 278.71      & 8.73    & 0.0                & 0.0    &        9.89 &  $1.2\times10^{-12} $     \\\hline
1E 1841-045           & 280.33      & 4.93    & 7.43               & 0.47  &       23.8  &$3.1\times10^{-12} $       \\\hline
3XMM J185246.6+003317 & 283.19      & 0.55    & 23.2             & 3.34  &    47.9   &$6.8\times10^{-13}$      \\\hline
SGR 1900+14           & 286.80     & 9.32   & 15.99              & 1.72   &      40.1  &$5.2\times10^{-12} $     \\\hline
SGR 1935+2154         & 293.73      & 21.89   & 17.29              & 2.47    &     43    &$5.5\times10^{-12} $  \\\hline
1E 2259+586           & 345.28      & 58.87  & 0.8                & 0.004       &    21.3  &$3.0\times10^{-12} $    \\\hline
SGR 0755-2933         & 118.92      & 29.53   & 17.06              & 2.69    &   39.1   &$5.1\times10^{-12} $   \\\hline

SGR 1801-23           & 270.24      & 22.92   &8.14               & 0.80      &    27.8  &$3.5\times10^{-12} $    \\\hline
SGR 1806-20           & 272.04     & 20.61   & 0.0                & 0.0     &       9.79   &$1.2\times10^{-12} $  \\\hline
AX J1818.8-1559       & 274.71      & 15.97   & 0.0                & 0.0     &     14.19   &$1.8\times10^{-12} $    \\\hline
AX J1845.0-0258       & 281.22      & 2.91   & 0.0                & 0.0      &      22.1 &$2.9\times10^{-12} $    \\\hline
SGR 2013+34           & 303.48      & 34.33   & 0.0                & 0.0    &           19.4&$2.5\times10^{-12} $\\\hline
PSR J1846-0258        & 281.60    & 2.95   & 2.42 & 0.02 & 28.5 &$3.8\times10^{-12} $ 
\\ \hline
GRB 231115A & 130.75&73.5&0.0&0.0&10.4&$1.5\times10^{-12} $
\\ \hline
GRB 200415A& 11.87& –25.02&0.0&0.0&17.9
&$2.5\times10^{-10}$\\ \hline
GRB 051103 &148.14 &68.84 &0.0 & 0.0&11.4&$1.7\times10^{-12} $ \\ \hline 
GRB 070201 &11.07 &42.3 & 0.0&0.0 & 13.2&$1.7\times10^{-12}$\\ \hline 
GRB 070222 & 205.53&- 26.87 &0.0 &0.0 &10 &$1.6\times10^{-10 }$\\ \hline 
GRB 180128A &12.3 &-26.1 &2.78& 0.046& 25.2&$3.6\times10^{-10} $\\ \hline 

\end{tabular}
 \caption{\label{table1} Results from spatial coincidence analysis between IceCube neutrinos and magnetars. The last three columns denote the number of observed signal events ($n_s$), the maximum observed significance ($TS_{max}$) as described in Eq.~\ref{eq:ts},  and  95\% c.l. upper limit on the number of signal muon-induced neutrino events from magnetars (column 6), respectively. The last column shows the differential neutrino flux limit evaluated at neutrino energy of 100 TeV for a spectral index of -2.53. The flux limit can be evaluated for any other neutrino energy according to Eq.~\ref{eq:phinu}.}
\end{table}
\end{center}

\bibliography{main}

@ARTICLE{Ioka05,
       author = {{Ioka}, Kunihito and {Razzaque}, Soebur and {Kobayashi}, Shiho and {M{\'e}sz{\'a}ros}, Peter},
        title = "{TeV-PeV Neutrinos from Giant Flares of Magnetars and the Case of SGR 1806-20}",
      journal = {\apj},
         year = 2005,
        month = nov,
       volume = {633},
       number = {2},
        pages = {1013-1017},
          doi = {10.1086/466514},
archivePrefix = {arXiv},
       eprint = {astro-ph/0503279},
 primaryClass = {astro-ph},
       adsurl = {https://ui.adsabs.harvard.edu/abs/2005ApJ...633.1013I}
}

@ARTICLE{Zhang03,
       author = {{Zhang}, Bing and {Dai}, Z.~G. and {M{\'e}sz{\'a}ros}, P. and {Waxman}, E. and {Harding}, A.~K.},
        title = "{High-Energy Neutrinos from Magnetars}",
      journal = {\apj},
         year = 2003,
        month = sep,
       volume = {595},
       number = {1},
        pages = {346-351},
          doi = {10.1086/377192},
archivePrefix = {arXiv},
       eprint = {astro-ph/0210382},
 primaryClass = {astro-ph},
       adsurl = {https://ui.adsabs.harvard.edu/abs/2003ApJ...595..346Z}
}

@ARTICLE{Woods03,
       author = {{Woods}, Peter M.},
        title = "{The dynamic behavior of soft gamma repeaters}",
      journal = {Advances in Space Research},
         year = 2004,
        month = jan,
       volume = {33},
       number = {4},
        pages = {630-637},
          doi = {10.1016/j.asr.2003.04.022},
archivePrefix = {arXiv},
       eprint = {astro-ph/0304372},
 primaryClass = {astro-ph},
       adsurl = {https://ui.adsabs.harvard.edu/abs/2004AdSpR..33..630W}
}

@ARTICLE{Luo05,
       author = {{Luo}, Qinghuan},
        title = "{High energy neutrinos from fast spinning magnetars}",
      journal = {Astroparticle Physics},
         year = 2005,
        month = dec,
       volume = {24},
       number = {4-5},
        pages = {301-315},
          doi = {10.1016/j.astropartphys.2005.07.003},
archivePrefix = {arXiv},
       eprint = {astro-ph/0507395},
 primaryClass = {astro-ph},
       adsurl = {https://ui.adsabs.harvard.edu/abs/2005APh....24..301L}
}

@ARTICLE{Dey16,
       author = {{Dey}, Rajat K. and {Ray}, Sabyasachi and {Dam}, Sandip},
        title = "{Searching for PeV neutrinos from photomeson interactions in magnetars}",
      journal = {EPL (Europhysics Letters)},
         year = 2016,
        month = sep,
       volume = {115},
       number = {6},
        pages = {69002},
          doi = {10.1209/0295-5075/115/69002},
archivePrefix = {arXiv},
       eprint = {1603.07833},
 primaryClass = {astro-ph.HE},
       adsurl = {https://ui.adsabs.harvard.edu/abs/2016EL....11569002D}
}

@ARTICLE{IceCubeGen2,
       author = {{Aartsen}, M.~G. and {Abbasi}, R. and {Ackermann}, M. and {Adams}, J. and {Aguilar}, J.~A. and {Ahlers}, M. and {Ahrens}, M. and {Alispach}, C. and {Allison}, P. and {Amin}, N.~M. and {Andeen}, K. and {Anderson}, T. and {Ansseau}, I. and {Anton}, G. and {Arg{\"u}elles}, C. and {Arlen}, T.~C. and {Auffenberg}, J. and {Axani}, S. and {Bagherpour}, H. and {Bai}, X. and {Balagopal V}, A. and {Barbano}, A. and {Bartos}, I. and {Bastian}, B. and {Basu}, V. and {Baum}, V. and {Baur}, S. and {Bay}, R. and {Beatty}, J.~J. and {Becker}, K. -H. and {Tjus}, J. Becker and {BenZvi}, S. and {Berley}, D. and {Bernardini}, E. and {Besson}, D.~Z. and {Binder}, G. and {Bindig}, D. and {Blaufuss}, E. and {Blot}, S. and {Bohm}, C. and {Bohmer}, M. and {B{\"o}ser}, S. and {Botner}, O. and {B{\"o}ttcher}, J. and {Bourbeau}, E. and {Bourbeau}, J. and {Bradascio}, F. and {Braun}, J. and {Bron}, S. and {Brostean-Kaiser}, J. and {Burgman}, A. and {Burley}, R.~T. and {Buscher}, J. and {Busse}, R.~S. and {Bustamante}, M. and {Campana}, M.~A. and {Carnie-Bronca}, E.~G. and {Carver}, T. and {Chen}, C. and {Chen}, P. and {Cheung}, E. and {Chirkin}, D. and {Choi}, S. and {Clark}, B.~A. and {Clark}, K. and {Classen}, L. and {Coleman}, A. and {Collin}, G.~H. and {Connolly}, A. and {Conrad}, J.~M. and {Coppin}, P. and {Correa}, P. and {Cowen}, D.~F. and {Cross}, R. and {Dave}, P. and {Deaconu}, C. and {De Clercq}, C. and {DeLaunay}, J.~J. and {De Kockere}, S. and {Dembinski}, H. and {Deoskar}, K. and {De Ridder}, S. and {Desai}, A. and {Desiati}, P. and {de Vries}, K.~D. and {de Wasseige}, G. and {de With}, M. and {DeYoung}, T. and {Dharani}, S. and {Diaz}, A. and {D{\'\i}az-V{\'e}lez}, J.~C. and {Dujmovic}, H. and {Dunkman}, M. and {DuVernois}, M.~A. and {Dvorak}, E. and {Ehrhardt}, T. and {Eller}, P. and {Engel}, R. and {Evans}, J.~J. and {Evenson}, P.~A. and {Fahey}, S. and {Farrag}, K. and {Fazely}, A.~R. and {Felde}, J. and {Fienberg}, A.~T. and {Filimonov}, K. and {Finley}, C. and {Fischer}, L. and {Fox}, D. and {Franckowiak}, A. and {Friedman}, E. and {Fritz}, A. and {Gaisser}, T.~K. and {Gallagher}, J. and {Ganster}, E. and {Garcia-Fernandez}, D. and {Garrappa}, S. and {Gartner}, A. and {Gerhard}, L. and {Gernhaeuser}, R. and {Ghadimi}, A. and {Glaser}, C. and {Glauch}, T. and {Gl{\"u}senkamp}, T. and {Goldschmidt}, A. and {Gonzalez}, J.~G. and {Goswami}, S. and {Grant}, D. and {Gr{\'e}goire}, T. and {Griffith}, Z. and {Griswold}, S. and {G{\"u}nd{\"u}z}, M. and {Haack}, C. and {Hallgren}, A. and {Halliday}, R. and {Halve}, L. and {Halzen}, F. and {Hanson}, J.~C. and {Hanson}, K. and {Hardin}, J. and {Haugen}, J. and {Haungs}, A. and {Hauser}, S. and {Hebecker}, D. and {Heinen}, D. and {Heix}, P. and {Helbing}, K. and {Hellauer}, R. and {Henningsen}, F. and {Hickford}, S. and {Hignight}, J. and {Hill}, C. and {Hill}, G.~C. and {Hoffman}, K.~D. and {Hoffmann}, B. and {Hoffmann}, R. and {Hoinka}, T. and {Hokanson-Fasig}, B. and {Holzapfel}, K. and {Hoshina}, K. and {Huang}, F. and {Huber}, M. and {Huber}, T. and {Huege}, T. and {Hughes}, K. and {Hultqvist}, K. and {H{\"u}nnefeld}, M. and {Hussain}, R. and {In}, S. and {Iovine}, N. and {Ishihara}, A. and {Jansson}, M. and {Japaridze}, G.~S. and {Jeong}, M. and {Jones}, B.~J.~P. and {Jonske}, F. and {Joppe}, R. and {Kalekin}, O. and {Kang}, D. and {Kang}, W. and {Kang}, X. and {Kappes}, A. and {Kappesser}, D. and {Karg}, T. and {Karl}, M. and {Karle}, A. and {Katori}, T. and {Katz}, U. and {Kauer}, M. and {Keivani}, A. and {Kellermann}, M. and {Kelley}, J.~L. and {Kheirandish}, A. and {Kim}, J. and {Kin}, K. and {Kintscher}, T. and {Kiryluk}, J. and {Kittler}, T. and {Kleifges}, M. and {Klein}, S.~R.},
        title = "{IceCube-Gen2: the window to the extreme Universe}",
      journal = {Journal of Physics G Nuclear Physics},
         year = 2021,
        month = jun,
       volume = {48},
       number = {6},
          eid = {060501},
        pages = {060501},
          doi = {10.1088/1361-6471/abbd48},
archivePrefix = {arXiv},
       eprint = {2008.04323},
 primaryClass = {astro-ph.HE},
       adsurl = {https://ui.adsabs.harvard.edu/abs/2021JPhG...48f0501A}
}

@ARTICLE{Murase09,
       author = {{Murase}, Kohta and {M{\'e}sz{\'a}ros}, Peter and {Zhang}, Bing},
        title = "{Probing the birth of fast rotating magnetars through high-energy neutrinos}",
      journal = {\prd},
         year = 2009,
        month = may,
       volume = {79},
       number = {10},
          eid = {103001},
        pages = {103001},
          doi = {10.1103/PhysRevD.79.103001},
archivePrefix = {arXiv},
       eprint = {0904.2509},
 primaryClass = {astro-ph.HE},
       adsurl = {https://ui.adsabs.harvard.edu/abs/2009PhRvD..79j3001M}
}

@ARTICLE{Ganguly17,
       author = {{Ganguly}, Shalini and {Desai}, Shantanu},
        title = "{Statistical significance of spectral lag transition in GRB 160625B}",
      journal = {Astroparticle Physics},
         year = 2017,
        month = sep,
       volume = {94},
        pages = {17-21},
          doi = {10.1016/j.astropartphys.2017.07.003},
archivePrefix = {arXiv},
       eprint = {1706.01202},
 primaryClass = {astro-ph.IM},
       adsurl = {https://ui.adsabs.harvard.edu/abs/2017APh....94...17G}
}

@PHDTHESIS{Desai04,
       author = {{Desai}, Shantanu},
        title = "{High energy neutrino astrophysics with Super-Kamiokande}",
       school = {Boston University, Massachusetts},
         year = 2004,
        month = may,
       adsurl = {https://ui.adsabs.harvard.edu/abs/2004PhDT.......222D}
}

@ARTICLE{Negro24,
       author = {{Negro}, Michela and {Younes}, George and {Wadiasingh}, Zorawar and {Burns}, Eric and {Trigg}, Aaron and {Baring}, Matthew},
        title = "{The role of magnetar transient activity in time-domain and multimessenger astronomy}",
      journal = {Frontiers in Astronomy and Space Sciences},
         year = 2024,
        month = jul,
       volume = {11},
          eid = {1388953},
        pages = {1388953},
          doi = {10.3389/fspas.2024.1388953},
archivePrefix = {arXiv},
       eprint = {2406.04967},
 primaryClass = {astro-ph.HE},
       adsurl = {https://ui.adsabs.harvard.edu/abs/2024FrASS..1188953N}
}

@ARTICLE{Halzen05,
       author = {{Halzen}, Francis and {Landsman}, Hagar and {Montaruli}, Teresa},
        title = "{TeV photons and Neutrinos from giant soft-gamma repeaters flares}",
      journal = {arXiv e-prints},
         year = 2005,
        month = mar,
          eid = {astro-ph/0503348},
        pages = {astro-ph/0503348},
          doi = {10.48550/arXiv.astro-ph/0503348},
archivePrefix = {arXiv},
       eprint = {astro-ph/0503348},
 primaryClass = {astro-ph},
       adsurl = {https://ui.adsabs.harvard.edu/abs/2005astro.ph..3348H}
}

@INCOLLECTION{Woods,
       author = {{Woods}, P.~M. and {Thompson}, C.},
        title = "{Soft gamma repeaters and anomalous X-ray pulsars: magnetar candidates}",
    booktitle = {Compact stellar X-ray sources},
         year = 2006,
       editor = {{Lewin}, Walter H.~G. and {van der Klis}, Michiel},
       volume = {39},
        pages = {547-586},
          doi = {10.48550/arXiv.astro-ph/0406133},
       adsurl = {https://ui.adsabs.harvard.edu/abs/2006csxs.book..547W}
}

@ARTICLE{Mereghetti,
       author = {{Mereghetti}, Sandro},
        title = "{The strongest cosmic magnets: soft gamma-ray repeaters and anomalous X-ray pulsars}",
      journal = {\aapr},
         year = 2008,
        month = jul,
       volume = {15},
       number = {4},
        pages = {225-287},
          doi = {10.1007/s00159-008-0011-z},
archivePrefix = {arXiv},
       eprint = {0804.0250},
 primaryClass = {astro-ph},
       adsurl = {https://ui.adsabs.harvard.edu/abs/2008A&ARv..15..225M}
}

@ARTICLE{Paczynski,
       author = {{Paczynski}, Bohdan},
        title = "{GB 790305 as a Very Strongly Magnetized Neutron Star}",
      journal = {\actaa},
         year = 1992,
        month = jul,
       volume = {42},
        pages = {145-153},
       adsurl = {https://ui.adsabs.harvard.edu/abs/1992AcA....42..145P}
}

@ARTICLE{DuncanThompson,
       author = {{Duncan}, Robert C. and {Thompson}, Christopher},
        title = "{Formation of Very Strongly Magnetized Neutron Stars: Implications for Gamma-Ray Bursts}",
      journal = {\apjl},
         year = 1992,
        month = jun,
       volume = {392},
        pages = {L9},
          doi = {10.1086/186413},
       adsurl = {https://ui.adsabs.harvard.edu/abs/1992ApJ...392L...9D}
}

@ARTICLE{ThompsonDuncan,
       author = {{Thompson}, Christopher and {Duncan}, Robert C.},
        title = "{The soft gamma repeaters as very strongly magnetized neutron stars - I. Radiative mechanism for outbursts}",
      journal = {\mnras},
         year = 1995,
        month = jul,
       volume = {275},
       number = {2},
        pages = {255-300},
          doi = {10.1093/mnras/275.2.255},
       adsurl = {https://ui.adsabs.harvard.edu/abs/1995MNRAS.275..255T}
}

@ARTICLE{Kaspirev,
       author = {{Kaspi}, Victoria M. and {Beloborodov}, Andrei M.},
        title = "{Magnetars}",
      journal = {\araa},
         year = 2017,
        month = aug,
       volume = {55},
       number = {1},
        pages = {261-301},
          doi = {10.1146/annurev-astro-081915-023329},
archivePrefix = {arXiv},
       eprint = {1703.00068},
 primaryClass = {astro-ph.HE},
       adsurl = {https://ui.adsabs.harvard.edu/abs/2017ARA&A..55..261K}
}

@INPROCEEDINGS{Espositorev,
       author = {{Esposito}, Paolo and {Rea}, Nanda and {Israel}, Gian Luca},
        title = "{Magnetars: A Short Review and Some Sparse Considerations}",
    booktitle = {Timing Neutron Stars: Pulsations, Oscillations and Explosions},
         year = 2021,
       editor = {{Belloni}, Tomaso M. and {M{\'e}ndez}, Mariano and {Zhang}, Chengmin},
       series = {Astrophysics and Space Science Library},
       volume = {461},
        month = jan,
        pages = {97-142},
          doi = {10.1007/978-3-662-62110-3_3},
archivePrefix = {arXiv},
       eprint = {1803.05716},
 primaryClass = {astro-ph.HE},
       adsurl = {https://ui.adsabs.harvard.edu/abs/2021ASSL..461...97E}
}

@ARTICLE{Esposito,
       author = {{Coti Zelati}, Francesco and {Rea}, Nanda and {Pons}, Jos{\'e} A. and {Campana}, Sergio and {Esposito}, Paolo},
        title = "{Systematic study of magnetar outbursts}",
      journal = {\mnras},
         year = 2018,
        month = feb,
       volume = {474},
       number = {1},
        pages = {961-1017},
          doi = {10.1093/mnras/stx2679},
archivePrefix = {arXiv},
       eprint = {1710.04671},
 primaryClass = {astro-ph.HE},
       adsurl = {https://ui.adsabs.harvard.edu/abs/2018MNRAS.474..961C}
}

@ARTICLE{Vyaas,
       author = {{Ramakrishnan}, Vyaas and {Desai}, Shantanu},
        title = "{Search for transient gamma-ray emission from magnetar flares using Fermi-LAT}",
      journal = {\jcap},
         year = 2025,
        month = jul,
       volume = {2025},
       number = {7},
          eid = {050},
        pages = {050},
          doi = {10.1088/1475-7516/2025/07/050},
archivePrefix = {arXiv},
       eprint = {2412.03900},
 primaryClass = {astro-ph.HE},
       adsurl = {https://ui.adsabs.harvard.edu/abs/2025JCAP...07..050R}
}

@ARTICLE{Halzen23,
       author = {{Halzen}, Francis},
        title = "{IceCube: Neutrinos from Active Galaxies}",
      journal = {arXiv e-prints},
         year = 2023,
        month = may,
          eid = {arXiv:2305.07086},
        pages = {arXiv:2305.07086},
          doi = {10.48550/arXiv.2305.07086},
archivePrefix = {arXiv},
       eprint = {2305.07086},
 primaryClass = {astro-ph.HE},
       adsurl = {https://ui.adsabs.harvard.edu/abs/2023arXiv230507086H}
}

@ARTICLE{Beacom,
       author = {{Zhou}, Bei and {Beacom}, John F.},
        title = "{Dimuons in neutrino telescopes: New predictions and first search in IceCube}",
      journal = {\prd},
         year = 2022,
        month = may,
       volume = {105},
       number = {9},
          eid = {093005},
        pages = {093005},
          doi = {10.1103/PhysRevD.105.093005},
archivePrefix = {arXiv},
       eprint = {2110.02974},
 primaryClass = {hep-ph},
       adsurl = {https://ui.adsabs.harvard.edu/abs/2022PhRvD.105i3005Z}
}

@ARTICLE{Pasumarti,
       author = {{Pasumarti}, Vibhavasu and {Desai}, Shantanu},
        title = "{Search for spatial coincidence between IceCube neutrinos and radio pulsars}",
      journal = {\jcap},
         year = 2022,
        month = dec,
       volume = {2022},
       number = {12},
          eid = {002},
        pages = {002},
          doi = {10.1088/1475-7516/2022/12/002},
archivePrefix = {arXiv},
       eprint = {2210.12804},
 primaryClass = {astro-ph.HE},
       adsurl = {https://ui.adsabs.harvard.edu/abs/2022JCAP...12..002P}
}

@ARTICLE{Pasumarti2,
       author = {{Pasumarti}, Vibhavasu and {Desai}, Shantanu},
        title = "{A stacked search for spatial coincidences between IceCube neutrinos and radio pulsars}",
      journal = {\jcap},
         year = 2024,
        month = apr,
       volume = {2024},
       number = {4},
          eid = {010},
        pages = {010},
          doi = {10.1088/1475-7516/2024/04/010},
archivePrefix = {arXiv},
       eprint = {2306.03427},
 primaryClass = {astro-ph.HE},
       adsurl = {https://ui.adsabs.harvard.edu/abs/2024JCAP...04..010P}
}

@article{Science,
    author = "Abbasi, R. and others",
    collaboration = "IceCube",
    title = "{Observation of high-energy neutrinos from the Galactic plane}",
    eprint = "2307.04427",
    archivePrefix = "arXiv",
    primaryClass = "astro-ph.HE",
    doi = "10.1126/science.adc9818",
    journal = "Science",
    volume = "380",
    number = "6652",
    pages = "adc9818",
    year = "2023"
}

@article{IceCubePWN,
    author = "Aartsen, M. G. and others",
    collaboration = "IceCube",
    title = "{IceCube Search for High-Energy Neutrino Emission from TeV Pulsar Wind Nebulae}",
    eprint = "2003.12071",
    archivePrefix = "arXiv",
    primaryClass = "astro-ph.HE",
    doi = "10.3847/1538-4357/ab9fa0",
    journal = "Astrophys. J.",
    volume = "898",
    number = "2",
    pages = "117",
    year = "2020"
}

@INPROCEEDINGS{Rea11,
       author = {{Rea}, Nanda and {Esposito}, Paolo},
        title = "{Magnetar outbursts: an observational review}",
    booktitle = {High-Energy Emission from Pulsars and their Systems},
         year = 2011,
       editor = {{Torres}, Diego F. and {Rea}, Nanda},
       series = {Astrophysics and Space Science Proceedings},
       volume = {21},
        month = jan,
        pages = {247},
          doi = {10.1007/978-3-642-17251-9_21},
archivePrefix = {arXiv},
       eprint = {1101.4472},
 primaryClass = {astro-ph.GA},
       adsurl = {https://ui.adsabs.harvard.edu/abs/2011ASSP...21..247R}
}

@ARTICLE{Rott,
       author = {{Kumar}, Jason and {Rott}, Carsten and {Sandick}, Pearl and {Tapia-Arellano}, Natalia},
        title = "{Are there correlations in the HAWC and IceCube high energy skymaps outside the Galactic plane?}",
      journal = {\prd},
         year = 2024,
        month = jul,
       volume = {110},
       number = {2},
          eid = {023009},
        pages = {023009},
          doi = {10.1103/PhysRevD.110.023009},
archivePrefix = {arXiv},
       eprint = {2312.15125},
 primaryClass = {hep-ph},
       adsurl = {https://ui.adsabs.harvard.edu/abs/2024PhRvD.110b3009K}
}

@ARTICLE{Gavri08,
       author = {{Gavriil}, F.~P. and {Gonzalez}, M.~E. and {Gotthelf}, E.~V. and {Kaspi}, V.~M. and {Livingstone}, M.~A. and {Woods}, P.~M.},
        title = "{Magnetar-Like Emission from the Young Pulsar in Kes 75}",
      journal = {Science},
         year = 2008,
        month = mar,
       volume = {319},
       number = {5871},
        pages = {1802},
          doi = {10.1126/science.1153465},
archivePrefix = {arXiv},
       eprint = {0802.1704},
 primaryClass = {astro-ph},
       adsurl = {https://ui.adsabs.harvard.edu/abs/2008Sci...319.1802G}
}

@ARTICLE{sknuastro,
       author = {{Abe}, K. and {Hosaka}, J. and {Iida}, T. and {Ishihara}, K. and {Kameda}, J. and {Koshio}, Y. and {Minamino}, A. and {Mitsuda}, C. and {Miura}, M. and {Moriyama}, S. and {Nakahata}, M. and {Obayashi}, Y. and {Ogawa}, H. and {Shiozawa}, M. and {Suzuki}, Y. and {Takeda}, A. and {Takeuchi}, Y. and {Higuchi}, I. and {Ishihara}, C. and {Ishitsuka}, M. and {Kajita}, T. and {Kaneyuki}, K. and {Mitsuka}, G. and {Nakayama}, S. and {Nishino}, H. and {Okada}, A. and {Okumura}, K. and {Saji}, C. and {Takenaga}, Y. and {Clark}, S. and {Desai}, S. and {Dufour}, F. and {Kearns}, E. and {Likhoded}, S. and {Litos}, M. and {Raaf}, J.~L. and {Stone}, J.~L. and {Sulak}, L.~R. and {Wang}, W. and {Goldhaber}, M. and {Casper}, D. and {Cravens}, J.~P. and {Dunmore}, J. and {Kropp}, W.~R. and {Liu}, D.~W. and {Mine}, S. and {Regis}, C. and {Smy}, M.~B. and {Sobel}, H.~W. and {Vagins}, M.~R. and {Ganezer}, K.~S. and {Hill}, J. and {Keig}, W.~E. and {Jang}, J.~S. and {Kim}, J.~Y. and {Lim}, I.~T. and {Scholberg}, K. and {Tanimoto}, N. and {Walter}, C.~W. and {Wendell}, R. and {Ellsworth}, R.~W. and {Tasaka}, S. and {Guillian}, G. and {Learned}, J.~G. and {Matsuno}, S. and {Messier}, M.~D. and {Hayato}, Y. and {Ichikawa}, A.~K. and {Ishida}, T. and {Ishii}, T. and {Iwashita}, T. and {Kobayashi}, T. and {Nakadaira}, T. and {Nakamura}, K. and {Nitta}, K. and {Oyama}, Y. and {Totsuka}, Y. and {Suzuki}, A.~T. and {Hasegawa}, M. and {Hiraide}, K. and {Kato}, I. and {Maesaka}, H. and {Nakaya}, T. and {Nishikawa}, K. and {Sasaki}, T. and {Sato}, H. and {Yamamoto}, S. and {Yokoyama}, M. and {Haines}, T.~J. and {Dazeley}, S. and {Hatakeyama}, S. and {Svoboda}, R. and {Sullivan}, G.~W. and {Turcan}, D. and {Swanson}, M. and {Clough}, A. and {Habig}, A. and {Fukuda}, Y. and {Sato}, T. and {Itow}, Y. and {Koike}, T. and {Jung}, C.~K. and {Kato}, T. and {Kobayashi}, K. and {Malek}, M. and {McGrew}, C. and {Sarrat}, A. and {Terri}, R. and {Yanagisawa}, C. and {Tamura}, N. and {Sakuda}, M. and {Sugihara}, M. and {Kuno}, Y. and {Yoshida}, M. and {Kim}, S.~B. and {Yang}, B.~S. and {Yoo}, J. and {Ishizuka}, T. and {Okazawa}, H. and {Choi}, Y. and {Seo}, H.~K. and {Gando}, Y. and {Hasegawa}, T. and {Inoue}, K. and {Ishii}, H. and {Nishijima}, K. and {Ishino}, H. and {Watanabe}, Y. and {Koshiba}, M. and {Kielczewska}, D. and {Zalipska}, J. and {Berns}, H. and {Gran}, R. and {Shiraishi}, K.~K. and {Stachyra}, A. and {Thrane}, E. and {Washburn}, K. and {Wilkes}, R.~J. and {Super-KAMIOKANDE Collaboration}},
        title = "{High-Energy Neutrino Astronomy Using Upward-going Muons in Super-Kamiokande I}",
      journal = {\apj},
         year = 2006,
        month = nov,
       volume = {652},
       number = {1},
        pages = {198-205},
          doi = {10.1086/508016},
archivePrefix = {arXiv},
       eprint = {astro-ph/0606413},
 primaryClass = {astro-ph},
       adsurl = {https://ui.adsabs.harvard.edu/abs/2006ApJ...652..198A}
}

@ARTICLE{Shifa,
       author = {{Shifa M}, Fathima and {Desai}, Shantanu},
        title = "{Search for spatial coincidence between IceCube neutrinos and gamma-ray bright red dwarfs}",
      journal = {Journal of High Energy Astrophysics},
         year = 2025,
        month = jul,
       volume = {47},
          eid = {100366},
        pages = {100366},
          doi = {10.1016/j.jheap.2025.100366},
archivePrefix = {arXiv},
       eprint = {2410.16394},
 primaryClass = {astro-ph.HE},
       adsurl = {https://ui.adsabs.harvard.edu/abs/2025JHEAp..4700366S}
}

@ARTICLE{skshowering,
       author = {{Desai}, S. and {Abe}, K. and {Hayato}, Y. and {Iida}, K. and {Ishihara}, K. and {Kameda}, J. and {Koshio}, Y. and {Minamino}, A. and {Mitsuda}, C. and {Miura}, M. and {Moriyama}, S. and {Nakahata}, M. and {Obayashi}, Y. and {Ogawa}, H. and {Shiozawa}, M. and {Suzuki}, Y. and {Takeda}, A. and {Takeuchi}, Y. and {Ueshima}, K. and {Watanabe}, H. and {Yamada}, S. and {Higuchi}, I. and {Ishihara}, C. and {Ishitsuka}, M. and {Kajita}, T. and {Kaneyuki}, K. and {Mitsuka}, G. and {Nakayama}, S. and {Nishino}, H. and {Okumura}, K. and {Saji}, C. and {Takenaga}, Y. and {Clark}, S.~T. and {Dufour}, F. and {Kearns}, E. and {Likhoded}, S. and {Raaf}, J.~L. and {Stone}, J.~L. and {Sulak}, L.~R. and {Wang}, W. and {Goldhaber}, M. and {Casper}, D. and {Cravens}, J.~P. and {Dunmore}, J. and {Kropp}, W.~R. and {Liu}, D.~W. and {Mine}, S. and {Regis}, C. and {Smy}, M.~B. and {Sobel}, H.~W. and {Vagins}, M.~R. and {Ganezer}, K.~S. and {Hartfiel}, B. and {Hill}, J. and {Keig}, W.~E. and {Jang}, J.~S. and {Jeong}, I.~S. and {Kim}, J.~Y. and {Lim}, I.~T. and {Fechner}, M. and {Scholberg}, K. and {Tanimoto}, N. and {Walter}, C.~W. and {Wendell}, R. and {Tasaka}, S. and {Guillian}, G. and {Learned}, J.~G. and {Matsuno}, S. and {Messier}, M.~D. and {Ichikawa}, A.~K. and {Ishida}, T. and {Ishii}, T. and {Kobayashi}, T. and {Nakadaira}, T. and {Nakamura}, K. and {Nitta}, K. and {Oyama}, Y. and {Totsuka}, Y. and {Suzuki}, A.~T. and {Hasegawa}, M. and {Hiraide}, K. and {Kato}, I. and {Maesaka}, H. and {Nakaya}, T. and {Nishikawa}, K. and {Sasaki}, T. and {Sato}, H. and {Yamamoto}, S. and {Yokoyama}, M. and {Haines}, T.~J. and {Dazeley}, S. and {Hatakeyama}, S. and {Svoboda}, R. and {Swanson}, M. and {Clough}, A. and {Gran}, R. and {Habig}, A. and {Fukuda}, Y. and {Sato}, T. and {Itow}, Y. and {Koike}, T. and {Tanaka}, T. and {Jung}, C.~K. and {Kato}, T. and {Kobayashi}, K. and {McGrew}, C. and {Sarrat}, A. and {Terri}, R. and {Yanagisawa}, C. and {Tamura}, N. and {Idehara}, Y. and {Sakuda}, M. and {Sugihara}, M. and {Kuno}, Y. and {Yoshida}, M. and {Kim}, S.~B. and {Yang}, B.~S. and {Yoo}, J. and {Ishizuka}, T. and {Okazawa}, H. and {Choi}, Y. and {Seo}, H.~K. and {Gando}, Y. and {Hasegawa}, T. and {Inoue}, K. and {Furuse}, Y. and {Ishii}, H. and {Nishijima}, K. and {Ishino}, H. and {Watanabe}, Y. and {Koshiba}, M. and {Kielczewska}, D. and {Berns}, H. and {Shiraishi}, K.~K. and {Thrane}, E. and {Washburn}, K. and {Wilkes}, R.~J.},
        title = "{Study of TeV neutrinos with upward showering muons in Super-Kamiokande}",
      journal = {Astroparticle Physics},
         year = 2008,
        month = feb,
       volume = {29},
        pages = {42-54},
          doi = {10.1016/j.astropartphys.2007.11.005},
archivePrefix = {arXiv},
       eprint = {0711.0053},
 primaryClass = {hep-ex},
       adsurl = {https://ui.adsabs.harvard.edu/abs/2008APh....29...42D}
}

@ARTICLE{LuoZhang,
       author = {{Luo}, Jia-Wei and {Zhang}, Bing},
        title = "{Blazar-IceCube neutrino association revisited}",
      journal = {\prd},
         year = 2020,
        month = may,
       volume = {101},
       number = {10},
          eid = {103015},
        pages = {103015},
          doi = {10.1103/PhysRevD.101.103015},
archivePrefix = {arXiv},
       eprint = {2004.09686},
 primaryClass = {astro-ph.HE},
       adsurl = {https://ui.adsabs.harvard.edu/abs/2020PhRvD.101j3015L}
}

@ARTICLE{Zhang21,
       author = {{Luo}, Jia-Wei and {Zhang}, Bing},
        title = "{Time-integrated constraint on neutrino flux of CHIME fast radio burst sources with 10-yr IceCube point-source data}",
      journal = {\mnras},
         year = 2024,
        month = oct,
       volume = {534},
       number = {1},
        pages = {70-75},
          doi = {10.1093/mnras/stae2071},
archivePrefix = {arXiv},
       eprint = {2112.11375},
 primaryClass = {astro-ph.HE},
       adsurl = {https://ui.adsabs.harvard.edu/abs/2024MNRAS.534...70L}
}

@ARTICLE{Smith21,
       author = {{Smith}, Daniel and {Hooper}, Dan and {Vieregg}, Abigail},
        title = "{Revisiting AGN as the source of IceCube's diffuse neutrino flux}",
      journal = {\jcap},
         year = 2021,
        month = mar,
       volume = {2021},
       number = {3},
          eid = {031},
        pages = {031},
          doi = {10.1088/1475-7516/2021/03/031},
archivePrefix = {arXiv},
       eprint = {2007.12706},
 primaryClass = {astro-ph.HE},
       adsurl = {https://ui.adsabs.harvard.edu/abs/2021JCAP...03..031S}
}

@article{IceCubeGRB,
    author = "Abbasi, R. and others",
    collaboration = "IceCube, Fermi Gamma-ray Burst Monitor",
    title = "{Searches for Neutrinos from Gamma-Ray Bursts Using the IceCube Neutrino Observatory}",
    eprint = "2205.11410",
    archivePrefix = "arXiv",
    primaryClass = "astro-ph.HE",
    doi = "10.3847/1538-4357/ac9785",
    journal = "Astrophys. J.",
    volume = "939",
    number = "2",
    pages = "116",
    year = "2022"
}

@article{IceCubeblazars,
    author = "Abbasi, R. and others",
    collaboration = "IceCube",
    title = "{Search for Astrophysical Neutrinos from 1FLE Blazars with IceCube}",
    eprint = "2207.04946",
    archivePrefix = "arXiv",
    primaryClass = "astro-ph.HE",
    doi = "10.3847/1538-4357/ac8de4",
    journal = "Astrophys. J.",
    volume = "938",
    number = "1",
    pages = "38",
    year = "2022"
}

@ARTICLE{Desai23,
       author = {{Desai}, Shantanu},
        title = "{A test of spatial coincidence between CHIME FRBs and IceCube TeV energy neutrinos}",
      journal = {Journal of Physics G Nuclear Physics},
         year = 2023,
        month = jan,
       volume = {50},
       number = {1},
          eid = {015201},
        pages = {015201},
          doi = {10.1088/1361-6471/aca03b},
archivePrefix = {arXiv},
       eprint = {2112.13820},
 primaryClass = {astro-ph.HE},
       adsurl = {https://ui.adsabs.harvard.edu/abs/2023JPhG...50a5201D}
}

@ARTICLE{Laha,
       author = {{Bouri}, Subhadip and {Parashari}, Priyank and {Das}, Mousumi and {Laha}, Ranjan},
        title = "{First search for high-energy neutrino emission from galaxy mergers}",
      journal = {\prd},
         year = 2025,
        month = mar,
       volume = {111},
       number = {6},
          eid = {063059},
        pages = {063059},
          doi = {10.1103/PhysRevD.111.063059},
archivePrefix = {arXiv},
       eprint = {2404.06539},
 primaryClass = {astro-ph.HE},
       adsurl = {https://ui.adsabs.harvard.edu/abs/2025PhRvD.111f3059B}
}

@article{IceCube2MASS,
    author = "Aartsen, M. G. and others",
    collaboration = "IceCube",
    title = "{Constraints on neutrino emission from nearby galaxies using the 2MASS redshift survey and IceCube}",
    eprint = "1911.11809",
    archivePrefix = "arXiv",
    primaryClass = "astro-ph.HE",
    doi = "10.1088/1475-7516/2020/07/042",
    journal = "JCAP",
    volume = "07",
    pages = "042",
    year = "2020"
}

@ARTICLE{Fang24,
       author = {{Zhou}, Zhuoyang and {Cisewski-Kehe}, Jessi and {Fang}, Ke and {Banerjee}, Arka},
        title = "{High-energy Neutrino Source Cross-correlations with Nearest-neighbor Distributions}",
      journal = {\apj},
         year = 2025,
        month = feb,
       volume = {979},
       number = {2},
          eid = {194},
        pages = {194},
          doi = {10.3847/1538-4357/ad924c},
archivePrefix = {arXiv},
       eprint = {2406.00796},
 primaryClass = {astro-ph.HE},
       adsurl = {https://ui.adsabs.harvard.edu/abs/2025ApJ...979..194Z}
}

@ARTICLE{Kamionkowski,
       author = {{Zhou}, Bei and {Kamionkowski}, Marc and {Liang}, Yun-feng},
        title = "{Search for high-energy neutrino emission from radio-bright AGN}",
      journal = {\prd},
         year = 2021,
        month = jun,
       volume = {103},
       number = {12},
          eid = {123018},
        pages = {123018},
          doi = {10.1103/PhysRevD.103.123018},
archivePrefix = {arXiv},
       eprint = {2103.12813},
 primaryClass = {astro-ph.HE},
       adsurl = {https://ui.adsabs.harvard.edu/abs/2021PhRvD.103l3018Z}
}

@INPROCEEDINGS{Wolf,
       author = {{Wolf}, M.},
        title = "{SkyLLH - A generalized Python-based tool for log-likelihood analyses in multi-messenger astronomy}",
    booktitle = {36th International Cosmic Ray Conference (ICRC2019)},
         year = 2019,
       series = {International Cosmic Ray Conference},
       volume = {36},
        month = jul,
          eid = {1035},
        pages = {1035},
          doi = {10.22323/1.358.01035},
archivePrefix = {arXiv},
       eprint = {1908.05181},
 primaryClass = {astro-ph.IM},
       adsurl = {https://ui.adsabs.harvard.edu/abs/2019ICRC...36.1035W}
}

@ARTICLE{Hooper,
       author = {{Hooper}, Dan and {Linden}, Tim and {Vieregg}, Abby},
        title = "{Active galactic nuclei and the origin of IceCube's diffuse neutrino flux}",
      journal = {\jcap},
         year = 2019,
        month = feb,
       volume = {2019},
       number = {2},
          eid = {012},
        pages = {012},
          doi = {10.1088/1475-7516/2019/02/012},
archivePrefix = {arXiv},
       eprint = {1810.02823},
 primaryClass = {astro-ph.HE},
       adsurl = {https://ui.adsabs.harvard.edu/abs/2019JCAP...02..012H}
}

@ARTICLE{Montaruli,
       author = {{Braun}, Jim and {Dumm}, Jon and {De Palma}, Francesco and {Finley}, Chad and {Karle}, Albrecht and {Montaruli}, Teresa},
        title = "{Methods for point source analysis in high energy neutrino telescopes}",
      journal = {Astroparticle Physics},
         year = 2008,
        month = may,
       volume = {29},
       number = {4},
        pages = {299-305},
          doi = {10.1016/j.astropartphys.2008.02.007},
archivePrefix = {arXiv},
       eprint = {0801.1604},
 primaryClass = {astro-ph},
       adsurl = {https://ui.adsabs.harvard.edu/abs/2008APh....29..299B}
}

\end{document}